# Photometry and spectrophotometry of the Herbig Ae star RR Tauri

David Boyd

## Abstract

Multicolour broadband photometry and flux calibrated low resolution spectroscopy between 2014 and 2018 are used to investigate the behaviour of the highly variable Herbig Ae star RR Tauri. As the star experienced a deep Algol-like fade by a factor of over 10 in flux, we found the change in spectral continuum to be "grey" while Hα emission flux halved. We also observed the (V-Rc) colour index reddening as the star faded suggesting emission detectable in the Rc-band but beyond our spectral response. We confirm that the circumstellar reddening of RR Tau is consistent with an extinction ratio Rv = 5 and that its spectral type is close to A0. According to its luminosity and temperature RR Tau is located in the H-R diagram among other HAEBE stars contracting onto the Zero Age Main Sequence and has a mass between 4 and 5 $M_\odot$.

## What do we know about RR Tau?

The star now known as RR Tau was first discovered to be variable by Lydia Ceraski in 1900 on photographic plates taken at Moscow Observatory (1). Merrill and Burwell (2) reported the star as having a bright Hα emission line in objective prism photographs taken at Mt Wilson. Herbig (3) included RR Tau as one of 26 stars with spectral type Ae or Be associated with nebulosity (the 'e' denoting the presence of emission lines in their spectra) which he predicted were still contracting to the main sequence. These stars have come to be known as Herbig Ae/Be (or HAEBE) stars and have masses in the range 2 to 8 $M_\odot$. Herbig reported RR Tau as varying irregularly between 10.2 and 14.2. Herbig's conjecture that HAEBE stars were young was confirmed by Strom et al. (4) who showed that they are pre-main sequence stars surrounded by circumstellar dust.

Over the following years many authors discussed the nature of HAEBE stars, and RR Tau in particular, including among others Grinin (5), Catala (6), Grinin (7), Rostropchina et al. (8), Grinin (9), Rodgers et al. (10), Hernandez et al. (11), Mendigutia et al. (12) and references therein. A consensus emerged that these stars are surrounded by a protoplanetary disc seen almost edge-on containing opaque clouds which orbit the star within the disc. Variable obscuration by these clouds is believed to be the cause of the irregular changes and occasional deep Algol-like minima seen in their light curves. In addition, they emit a stellar wind which generates emission lines in a circumstellar shell of gas and dust. This shell contains larger dust grains than occur in the interstellar medium and strongly reddens light emerging from the star. According to this model the star's brightest state represents its intrinsic luminosity.

However not everyone followed this general consensus. Herbst & Shevchenko (13) argued that the brightness variations may be due to variable accretion from circumstellar material. In their model fades are caused by a drop in accretion rate so the true luminosity of the star is seen at minimum light. It may be that a combination of variable obscuration and variable accretion mechanisms will be necessary to fully explain the observed behaviour of RR Tau.

RR Tau is informally classified as an UXOR star, a term invented by Herbst et al. (14) to classify the group of young massive variable stars with strong Hα emission lines whose variability characteristics are similar to those of the prototype UX Ori. However, there is still no commonly accepted definition of UXOR stars reflecting the uncertainty about the physical cause of their irregular variability. In their 1999 paper Herbst & Shevchenko (13) state "It is probably not an exaggeration to say that the

HAEBE stars are the only remaining class of variable stars with a substantial membership whose basic variability mechanism is still unclear."

Shenavrin et al. (15) reported V-band observations of RR Tau by the All Sky Automated Survey (16) over 13 years beginning in 2002 which show several Algol-like deep minima. The BAAVSS database (17) shows RR Tau varying irregularly over a visual magnitude range of 10.2 to 14.5, equivalent to a factor of 50 in brightness, with occasional non-periodic Algol-like minima. The BAA Handbook for 2013 contains a short article on RR Tau written by Gary Poyner (18). There is also an AAVSO *Variable Star Of The Season* article on RR Tau (19). The Digitized Sky Survey (20) POSS2 Red image of the region around RR Tau in Figure 1 shows extensive nebulosity, material from which the star is likely to have formed within the last few million years.

**Photometric observations**

V and Rc filtered images of RR Tau were recorded using a 0.35m Schmidt-Cassegrain telescope on 53 nights during four observing seasons between Dec 2014 and March 2018. A small number of B filtered images were also obtained during 2018. All images were dark-subtracted and flat-fielded before magnitudes were determined by differential photometry with respect to an ensemble of three nearby comparison stars selected from the AAVSO chart for RR Tau (21). The uncertainties on B, V and Rc magnitudes were calculated from a combination of computed photometry measurement errors and errors given for the magnitudes of the comparison stars used. Mean uncertainties in B, V and Rc magnitudes are 0.011, 0.009 and 0.010 respectively. All magnitude measurements are available in the BAAVSS database.

Over this period RR Tau's V magnitude varied erratically between 11.12 and 13.91 as shown in Figure 2. Between 14 March and 12 April 2016 the star experienced a deep Algol-like fade when its magnitude dropped from V=11.16 to V=13.91, a factor of 12.6 in flux. This occurred immediately before the star was lost at the end of the 2016 observing season.

Figure 3 shows that the (V-Rc) colour index reddened as the star faded, changing from 0.35 at its brightest to 0.68 at its faintest. A weighted linear fit to the relation between (V-Rc) and V gives:

$$(V-Rc) = 0.115(3) * V - 0.93(3) \qquad (1)$$

**Spectroscopic observations**

We recorded low resolution (5Å) spectra of RR Tau between 4000Å and 7000Å on 14 nights with a LISA spectrograph on a 0.28m Schmidt-Cassegrain telescope. The times of these spectra are marked in Figure 2. Spectra were bias and dark subtracted, flat fielded and wavelength calibrated using the spectrograph's internal Ar-Ne lamp. They were then corrected for instrumental and atmospheric response using spectra of an A type star with a known spectral profile located at the same air mass as RR Tau and recorded immediately prior to the RR Tau spectra. Finally spectra were flux calibrated by scaling the V-band flux using a V magnitude measured concurrently with the spectra.

Figure 4 shows three spectra recorded on 10 February 2016 when the star was bright at V mag 11.30, on 22 March 2016 at V mag 11.67 as it started to fade and on 9 April 2016 when the star had faded to V mag 13.16. The upper diagram shows the reduction in absolute flux of the continuum by a factor 5.5 between February and April. A reduction in flux of the Hα emission line is also visible. In the lower diagram all three spectra have been normalised to unity at the continuum level around Hα showing that the change in the continuum as the star faded was essentially "grey".

To investigate this further, the mean continuum fluxes at 4500Å and around Hα in all spectra were converted to magnitudes, V(4500) and V(Hα) respectively. The variations of [V(4500)-V] and [V-V(Hα)] with V are shown in Figure 5. The slope of weighted linear fits to these plots for [V(4500)-V] is 0.019±0.054 and for [V-V(Hα)] is 0.015±0.040 confirming that there was no significant colour change in the spectral continuum as the star faded. We note that Rodgers et al. (10) found the temperature of the photosphere of RR Tau did not change significantly as the star faded, which is consistent with our observations.

The Hα 6563 and forbidden [O I] 6300 emission lines are produced by a stellar wind in circumstellar material and persisted in emission through changes in the star's brightness. However there were significant changes in other spectral lines. In the star's bright state the Hβ 4861, Hγ 4340 and Hδ 4102 lines were in absorption and there were also absorption lines of the Fe II (42) multiplet (4924, 5018, 5169), He I 5876, and the Na I D lines (5890, 5896). In its faint state Hβ changed from absorption to emission while Hγ and Hδ remained in absorption and the other lines become undetectable. According to Hernandez et al. (11), the absorption lines are also believed to originate in circumstellar material.

**The Hα emission line**

Figure 6 shows how our measurements of the equivalent width, continuum level and absolute flux of the Hα emission line in RR Tau varied with the star's V magnitude. The uncertainties in measuring equivalent width and continuum flux are both estimated to be 10% and these are propagated to the uncertainty in emission line flux. We find the Hα equivalent width increasing approximately quadratically by a factor of ~5, the continuum level at Hα decreasing exponentially by a factor of ~10 and as a consequence the Hα absolute flux showing a decreasing trend by a factor of ~2 as the star fades by 2.5 magnitudes from V=11.3 to V=13.8.

On the other hand, Rodgers et al. (10) and Mendigutia et al. (12) saw no dependence of Hα flux on V magnitude in RR Tau. However Rodgers et al. (10) had no magnitude measurements concurrent with their spectra so used interpolated magnitudes from databases to flux calibrate their spectra. Given the rapid, often daily, changes in the magnitude of RR Tau, this is intrinsically less accurate than flux calibrating spectra using concurrently measured V magnitudes. Mendigutia et al. (12) only observed a V magnitude range of ~0.6 mag which would be insufficient to reveal a clear trend in flux.

**Why is the star apparently reddening as it fades?**

There appears to be a contradiction between the photometric observation that the (V-Rc) colour index reddens as the star fades and the spectroscopic observation that the colour of the continuum does not changes as it fades. A partial answer may lie in the behaviour of the Hα emission line which contributes to the light recorded by the Rc filter but not the V filter. While flux in the continuum drops by a factor 5.5 as the star fades, the Hα emission line flux only reduces by a factor 2. Although this contributes to a brightening in Rc relative to V, closer examination shows that there is insufficient flux in the Hα line to cause the observed degree of brightening in Rc. Other possible causes may be either changes in the continuum level or unseen emission features beyond the red end of our spectra at 7000Å which contribute to the flux measured in the Rc filter whose transmission extends to 8500Å.

Catala (6) proposed that the smaller drop in Hα flux relative to the drop in the continuum light from the photosphere observed when the star fades can be explained on the grounds that, even when the photosphere of the star is partially obscured by a dust cloud, a greater proportion of the Hα emission from the stellar wind will still reaches us. This is sometimes referred to as "nature's coronograph".

**Circumstellar extinction and reddening**

Rv is the ratio of total visual extinction Av to selective extinction E(B-V). The average value in the interstellar medium is Rv = 3.1. From an analysis of the colours of 39 HAEBE stars including RR Tau, Hernandez et al. (11) found the most consistent match of the reddening experienced by these stars was with Rv = 5. It is therefore likely that this increased reddening beyond that expected from the interstellar medium takes place in the dusty environment around the star. According to Hernandez et al. (11), this larger value of Rv indicates that the dust grains in this environment are on average larger than those in the interstellar medium.

**Spectral type of RR Tau**

Determining the spectral type of pre-main sequence stars which experience significant reddening has proved difficult and resulted in large uncertainties on estimated spectral types. Over the years RR Tau has been classified as anywhere between B8 and A5 by different authors, largely because different methods were used for classification. However in recent studies results have become more consistent. Grinin et al. (9) gave the spectral type of RR Tau as A0III-IV, Mora et al. (22) gave A0IV with an uncertainty of ±5 spectral subtypes, while Hernandez et al. (11) gave the spectral type of the photosphere of RR Tau as A0 with an uncertainty of ±2 spectral subtypes. According to Gray & Corbally (23) determining the luminosity class of HAEBE stars is often problematic. The literature is generally silent on the intrinsic colour index and temperature corresponding to spectral type A0IV, although Boehm-Vitense (24) reports there are no systematic differences in this respect between luminosity classes IV and V for A-type stars. We will therefore assume the intrinsic colour index of RR Tau is the same as for an A0V star. Pecaut and Mamajek (25) give intrinsic colours for A0V stars (B-V)$_0$ = 0.000, (V-Rc)$_0$ = 0.001.

From equation (1) we get (V-Rc) = 0.370 when V = 11.30 which gives E(V-Rc) = (V-Rc) – (V-Rc)$_0$ = 0.369. Using the parameterised extinction relationship given in Cardelli et al. (26) with Rv = 5 we can derive the relation

$$E(B-V) = 1.5855 * E(V-Rc) \qquad (2)$$

This gives E(B-V) = 0.584 for RR Tau when V = 11.30. With Rv = 5, this corresponds to a total visual extinction Av = 2.92 mag. Since (B-V)$_0$ = 0.000, we have (B-V) = 0.584. This is consistent with the value of (B-V) = 0.581 which we obtain from our measurements of (B-V) when V = 11.30.

We can also use the above calculation to investigate the consequence of assuming different spectral types for RR Tau. Using the values of intrinsic colour for different spectral types in Pecaut and Mamajek (25), the following values of (B-V) were calculated for each spectral type assuming V=11.30.

   B9V    (B-V) = 0.560
   A0V    (B-V) = 0.584
   A1V    (B-V) = 0.599

Although not definitive evidence in favour of type A0, it is consistent with that assumption.

If we were to adopt a value of Rv = 3.1, the average value for the interstellar medium, eqn (2) becomes

$$E(B-V) = 2.0539 * E(V-Rc) \tag{3}$$

This would give E(B-V) = 0.760 and hence (B-V) = 0.760, which is clearly inconsistent with our (B-V) measurements. This confirms that the value of Rv = 5 found by Hernandez et al. (11) for HAEBE stars and attributed to increased extinction local to the star is likely to be valid for RR Tau.

According to the variable obscuration model for HAEBE stars, the brightest state of RR Tau represents its intrinsic luminosity. We therefore assume that our spectrum taken on 10 February 2016 with the star at V=11.30 is a heavily reddened version of its presumed A0 spectral type. Using the parameterised extinction relationship in Cardelli et al. (26) with E(B-V) = 0.584 to deredden the spectrum we get the spectral profile shown in Figure 7. Also shown is the library spectrum of an A0V star from the Pickles Stellar Spectral Flux Library (27). The match is acceptable within the ±2 spectral subtype uncertainty in the published spectral type of RR Tau.

**Other properties of RR Tau**

The range of possible spectral types for RR Tau in the literature is reflected in a range of possible effective temperatures. Grinin et al. (9) give 9750K, Rodgers et al. (10) give the range 9000 – 10,000K, Hernandez et al. (11) give 9770 K and Montesinos et al. (28) give 10,000 K. For consistency with the spectral type of A0V we adopted from Pecaut and Mamajek (25), we adopt their effective temperature Teff = 9700 K and therefore log (Teff) = 3.99.

Several wide-ranging values have been given for the distance of RR Tau. Strom et al. (4) give 800pc citing Herbig (3) as the source, Grinin et al. (9) give 380 pc citing Rostopchina (29). None of these values comes with an estimate of its uncertainty. Montesinos et al. (28) derive a much larger distance of 2103±540 pc, although they urge caution in using this figure. Gaia Data Release 2 (30) gives a distance of 763±27 pc which is better constrained than the above values and we adopt this value for the distance of RR Tau.

As the canonical value for interstellar extinction in the galactic disc is 1 mag/kpc, we would expect visual extinction to be less than 1 magnitude. As we observe extinction of almost 3 magnitudes, this supports the premise that there is substantial circumstellar extinction.

At galactic coordinates (l = 181°, b = -2°) and a distance of 763 pc, RR Tau does not lie in any of the major galactic molecular clouds listed in Schlafly et al. (31). Its position on the sky places it approximately 15° east of the Taurus Molecular Cloud. It is also much further away than the Taurus Molecular Cloud which Guedel et al. (32) locate at 140 pc. At that distance and direction RR Tau is probably located in the Orion Arm of the Milky Way (33).

At a distance d = 763 pc and with extinction Av = 2.92 and apparent V magnitude mv = 11.30, we can calculate its absolute magnitude Mv from the relation

$$Mv = mv - Av - 5\log(d/10) \tag{4}$$

This gives Mv = -1.03. According to Pecaut & Mamjek (25), the bolometric correction for an A0V star is -0.24 making the bolometric magnitude of RR Tau Mbol = -1.27. From this we can find the intrinsic luminosity of RR Tau from the equation

$$\log(L/L_\odot) = -0.4(Mbol - 4.75) \tag{5}$$

This gives log(L/L$_\odot$) = 2.41 and together with log (Teff) = 3.99, this locates RR Tau in the H-R diagram shown in Figure 8 which is adapted from Figure 10 of Hernandez et al. (11). The solid lines representing the evolutionary tracks of stars of different mass contracting to the Zero Age Main Sequence (marked ZAMS) suggest the mass of RR Tau is around 4-5 M$_\odot$.

The commonly adopted mass luminosity relationship for main sequence stars

$$(L/L_\odot) = (M/M_\odot)^{3.5} \tag{6}$$

gives a mass of 4.88 M$_\odot$ consistent with the above mass estimate.

**Conclusions**

RR Tau is a young pre-main sequence star embedded in a complex environment comprising a stellar wind which generates emission lines, a protoplanetary disc containing opaque clouds which introduce irregular and occasionally deep obscuration of the star, and a dusty circumstellar shell which reddens the emerging starlight. By a combination of filtered broadband photometry and concurrent flux-calibrated spectroscopy, we found the change in continuum flux of the star to be grey as the star faded by a factor of over 10. The Hα equivalent width increased, the Hα continuum reduced sharply and the Hα emission line flux halved as the V magnitude reduced by 2.5 magnitudes. Surprisingly, given the continuum change in our spectra was observed to be grey, the (V-Rc) colour index reddened as the star faded suggesting that the Rc filter was detecting emission beyond our spectral response. The reddening introduced by the dusty circumstellar shell is consistent with a ratio of total visual to selective extinction Rv = 5. Dereddening the measured spectrum of RR Tau confirmed its spectral type is consistent with A0. According to its luminosity and temperature, RR Tau is located in the H-R diagram among other HAEBE stars contracting onto the Zero Age Main Sequence and has a mass between 4 and 5 M$_\odot$.

**Acknowledgements**

We are grateful for the comments of our referees which have helped to improve the paper. We acknowledge the work of the BAA in providing an archive of variable star observations and the work of the AAVSO in providing comparison star charts and sequences. This work has made use of data from the European Space Agency (ESA) mission Gaia processed by the Gaia Data Processing and Analysis Consortium (DPAC).

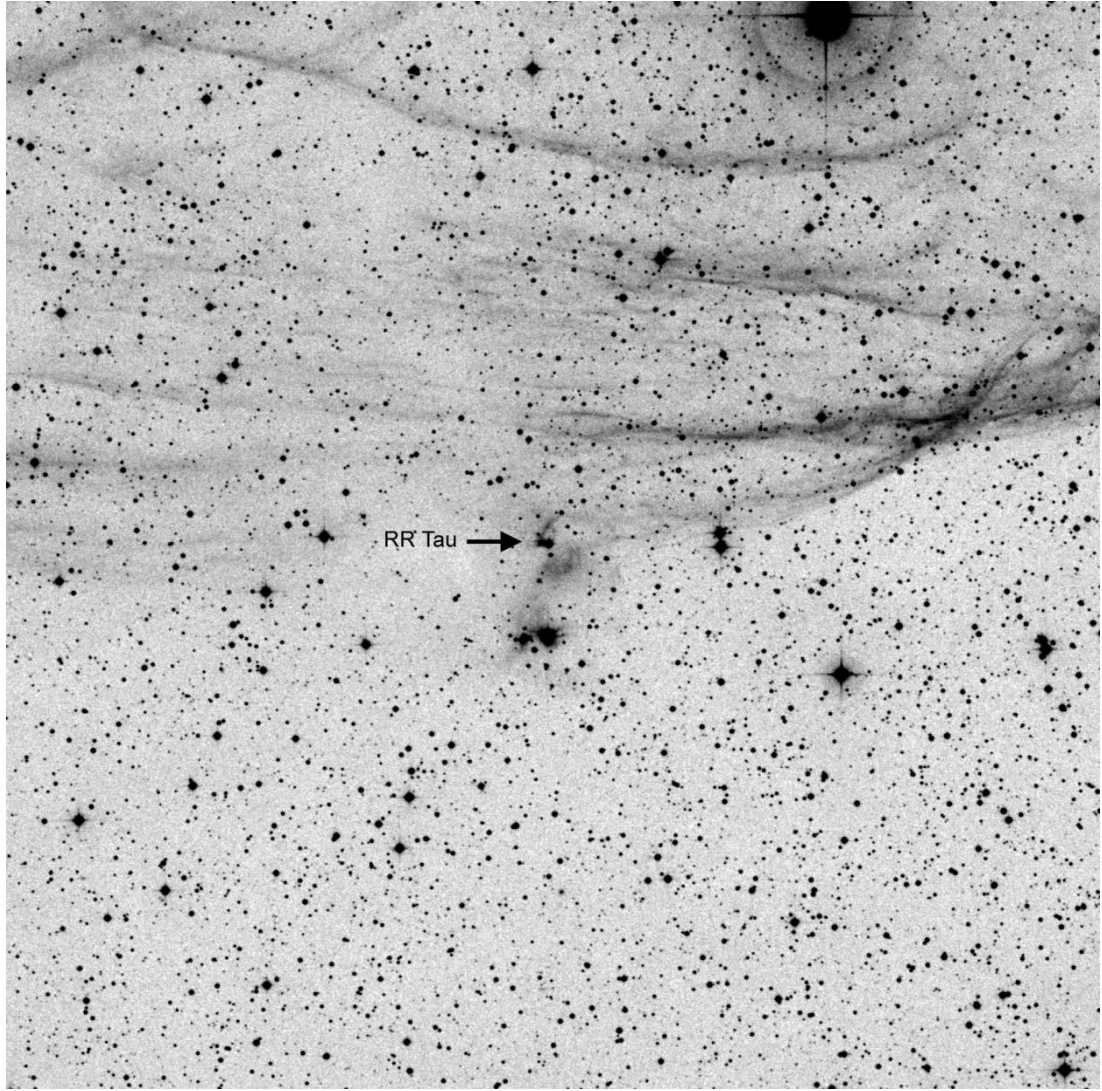

Figure 1. 30' x 30' region of the Digitized Sky Survey POSS2 Red image around RR Tau showing extensive nebulosity.

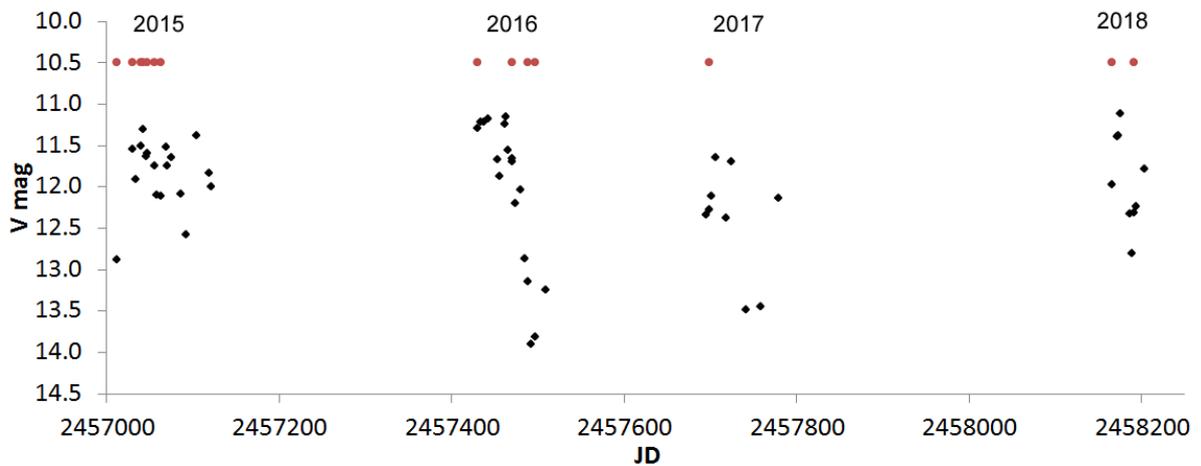

Figure 2. Measured V magnitudes of RR Tau between December 2014 and March 2018. Note the deep Algol-like minimum in April 2016. Also shown in red are the times at which spectra were recorded.

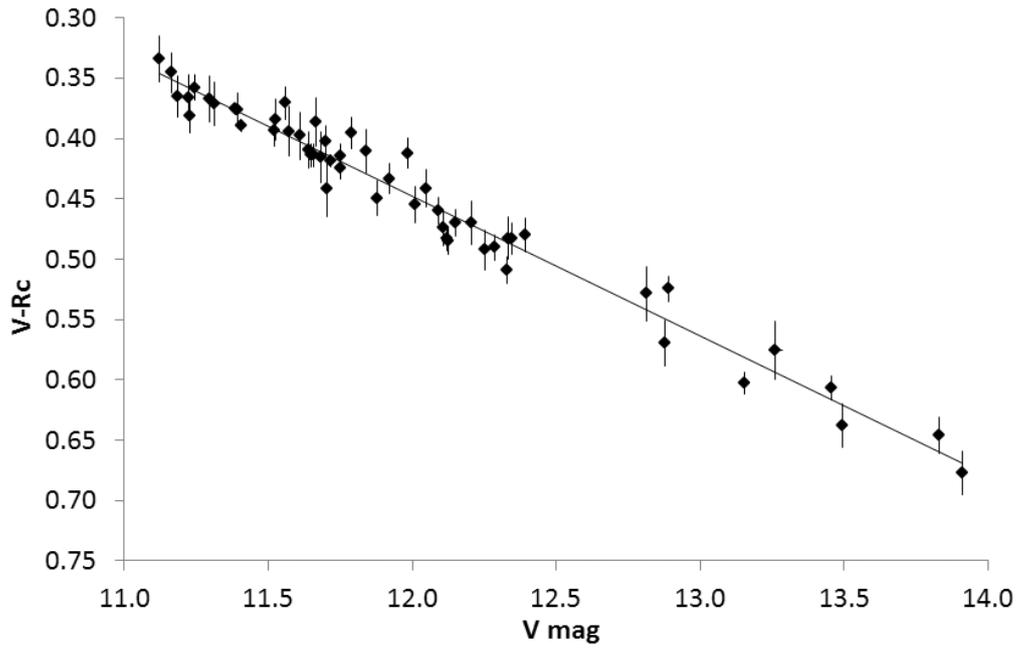

Figure 3. Variation of the (V-Rc) colour index of RR Tau with V magnitude. The line is a weighted linear fit. V magnitude errors in this and subsequent figures are smaller than the symbol.

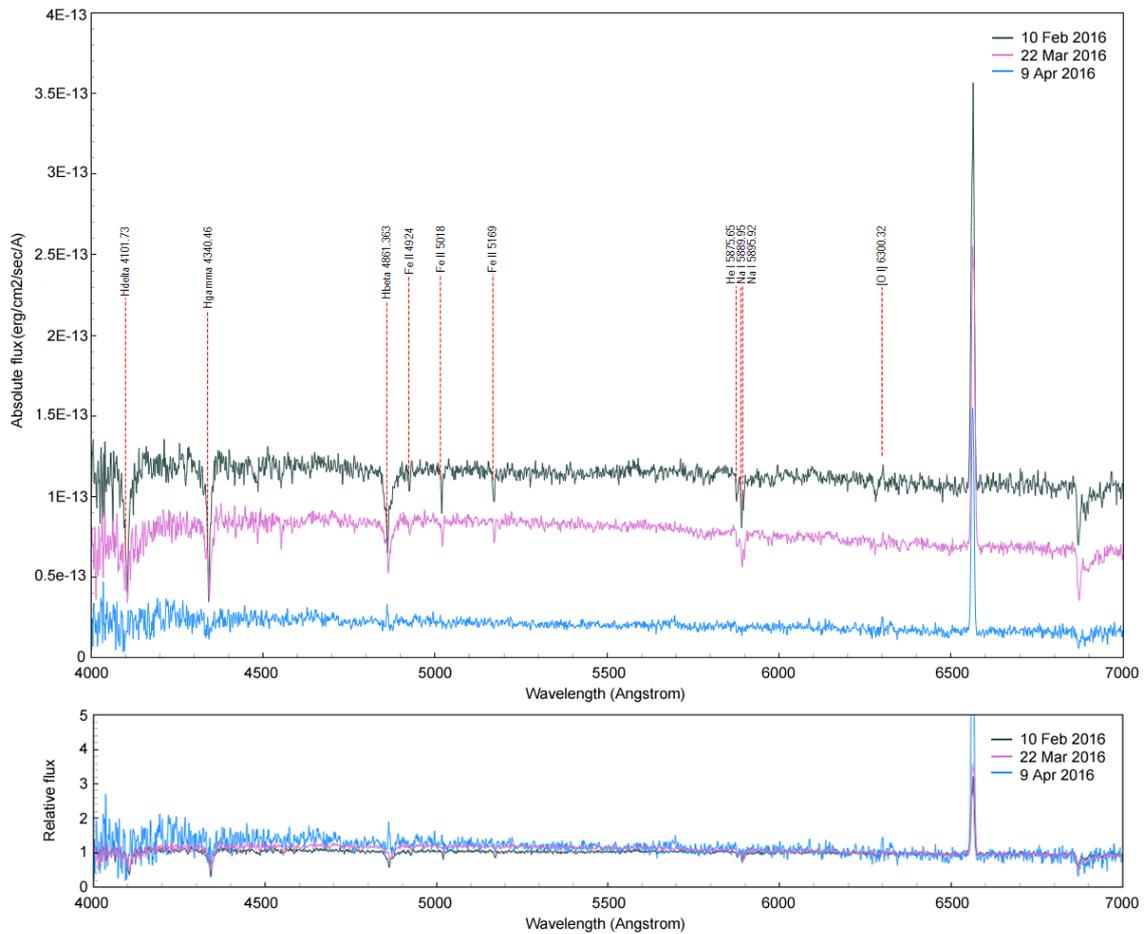

Figure 4. Spectra of RR Tau recorded on 10 February 2016 with the star at V = 11.30, on 22 March 2016 at V = 11.67 and on 9 April 2016 at V = 13.16. The upper diagram shows the reduction in absolute flux of the continuum by a factor 5.5 between February and April. In the lower diagram all three spectra have been normalised to unity at the continuum around Hα showing that the change in the continuum as the star faded was essentially "grey".

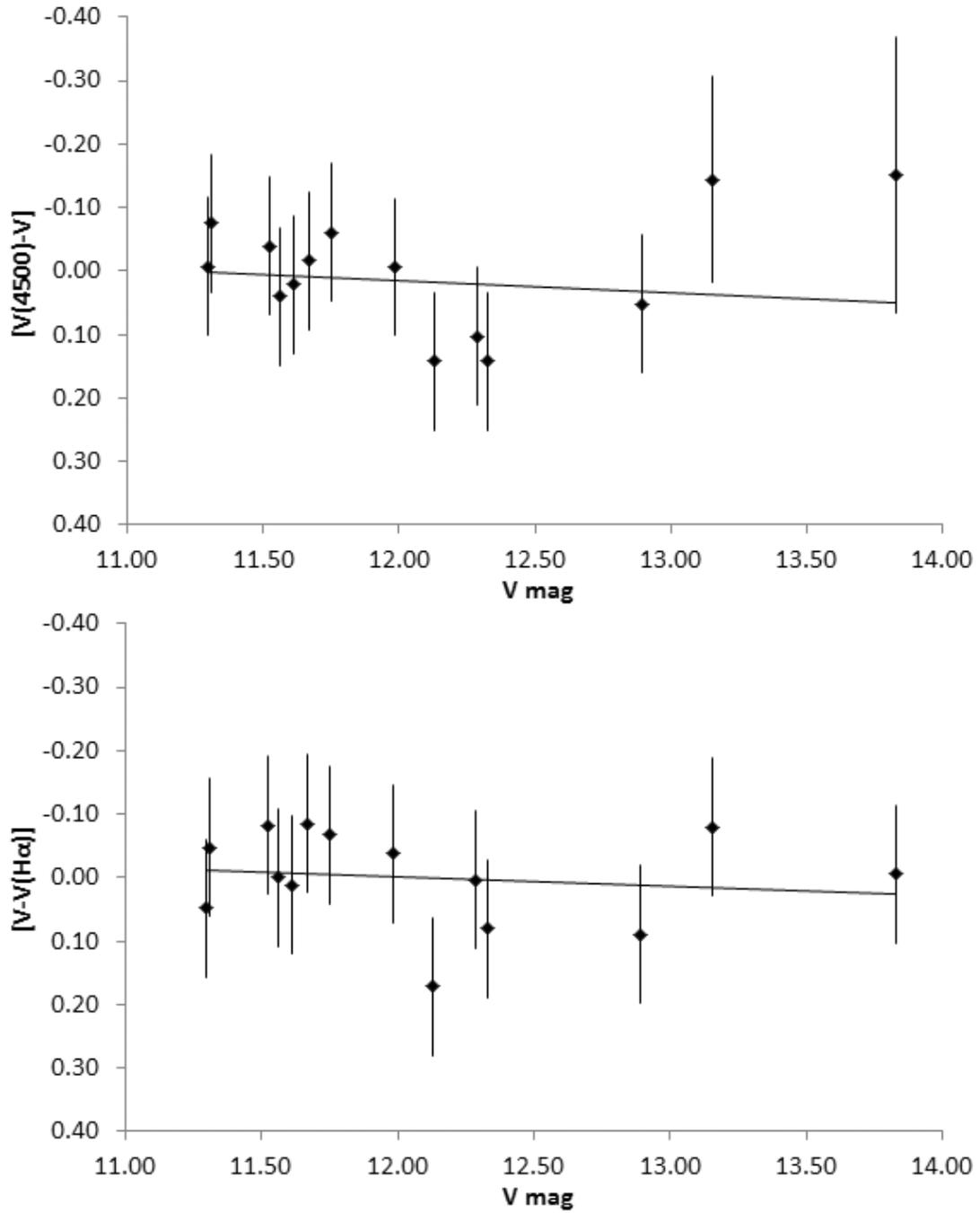

Figure 5. Variations of [V(4500)-V] and [V-V(Hα)] with V confirming that there was no significant colour change in the spectral continuum as the star faded. V(4500) and V(Hα) are respectively the mean continuum fluxes at 4500Å and around Hα converted to magnitudes.

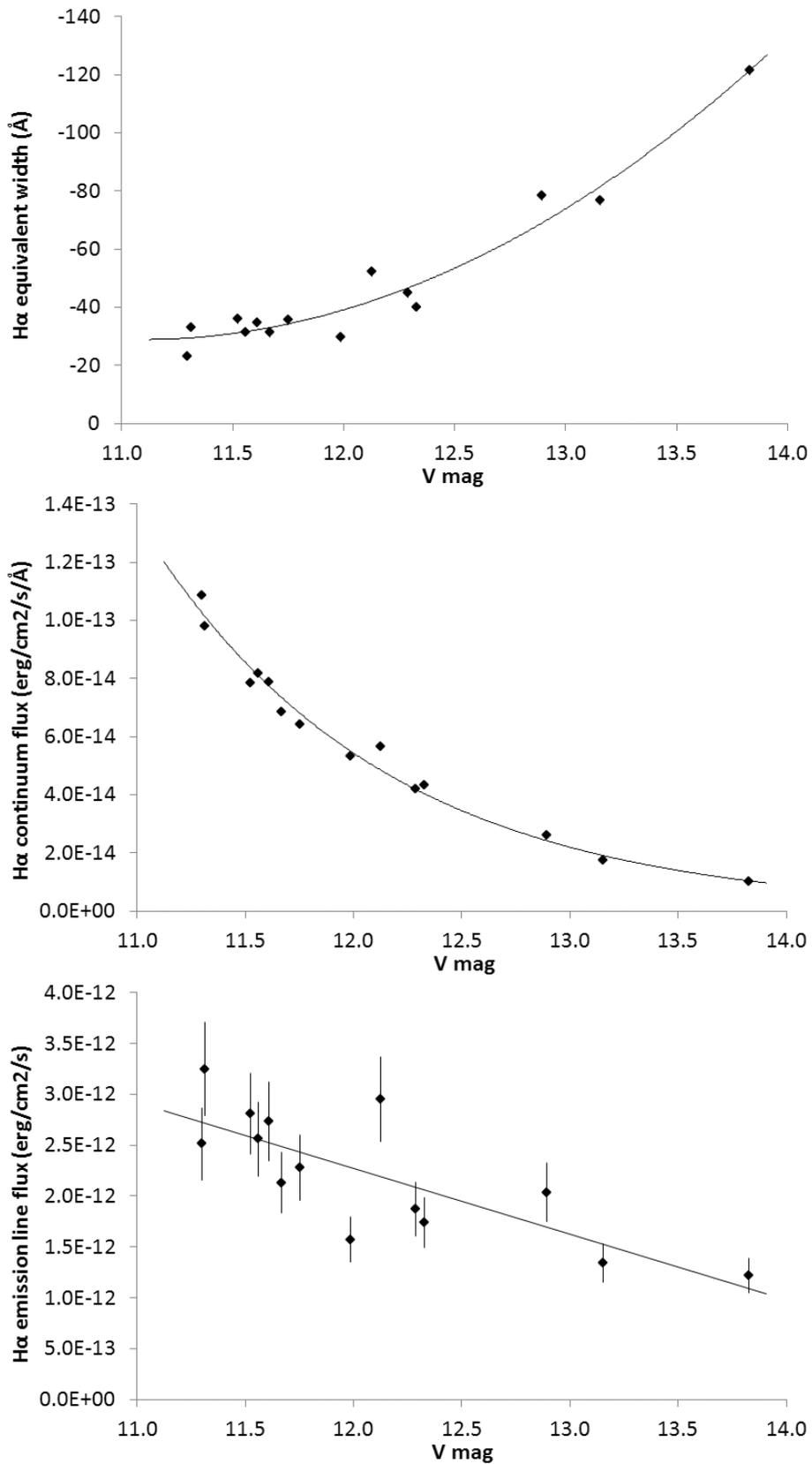

Figure 6. Variation with V magnitude of the Hα equivalent width (top) with a quadratic fit, Hα continuum flux (middle) with an exponential fit, and Hα emission line flux (bottom) with a linear fit. These fits are purely to guide the eye. The uncertainties in measuring equivalent width and continuum flux are both estimated to be 10% and these are propagated to the uncertainty in emission line flux.

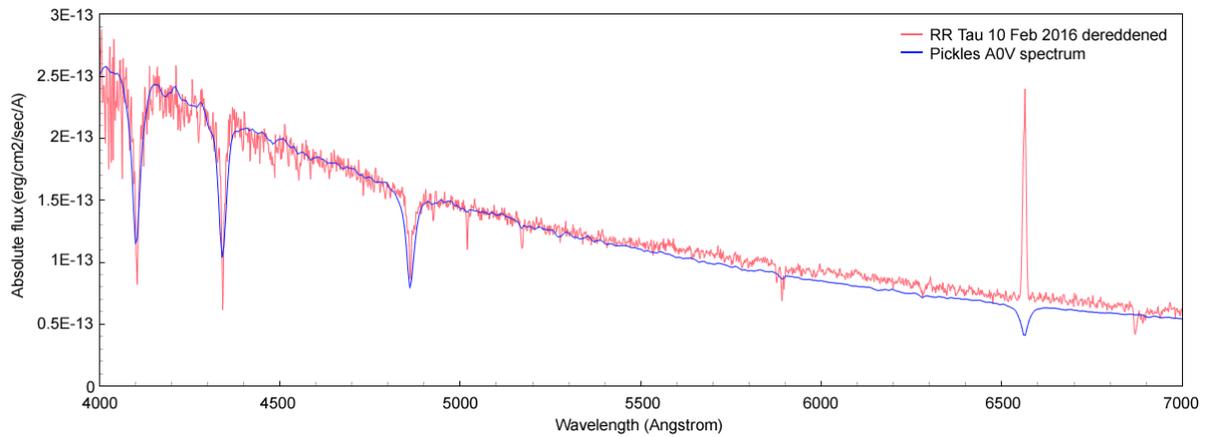

Figure 7. The spectral profile of RR Tau on 10 February 2016 dereddened with E(B-V) = 0.584. Also shown is the library spectrum of an A0V star from the Pickles Stellar Spectral Flux Library (27).

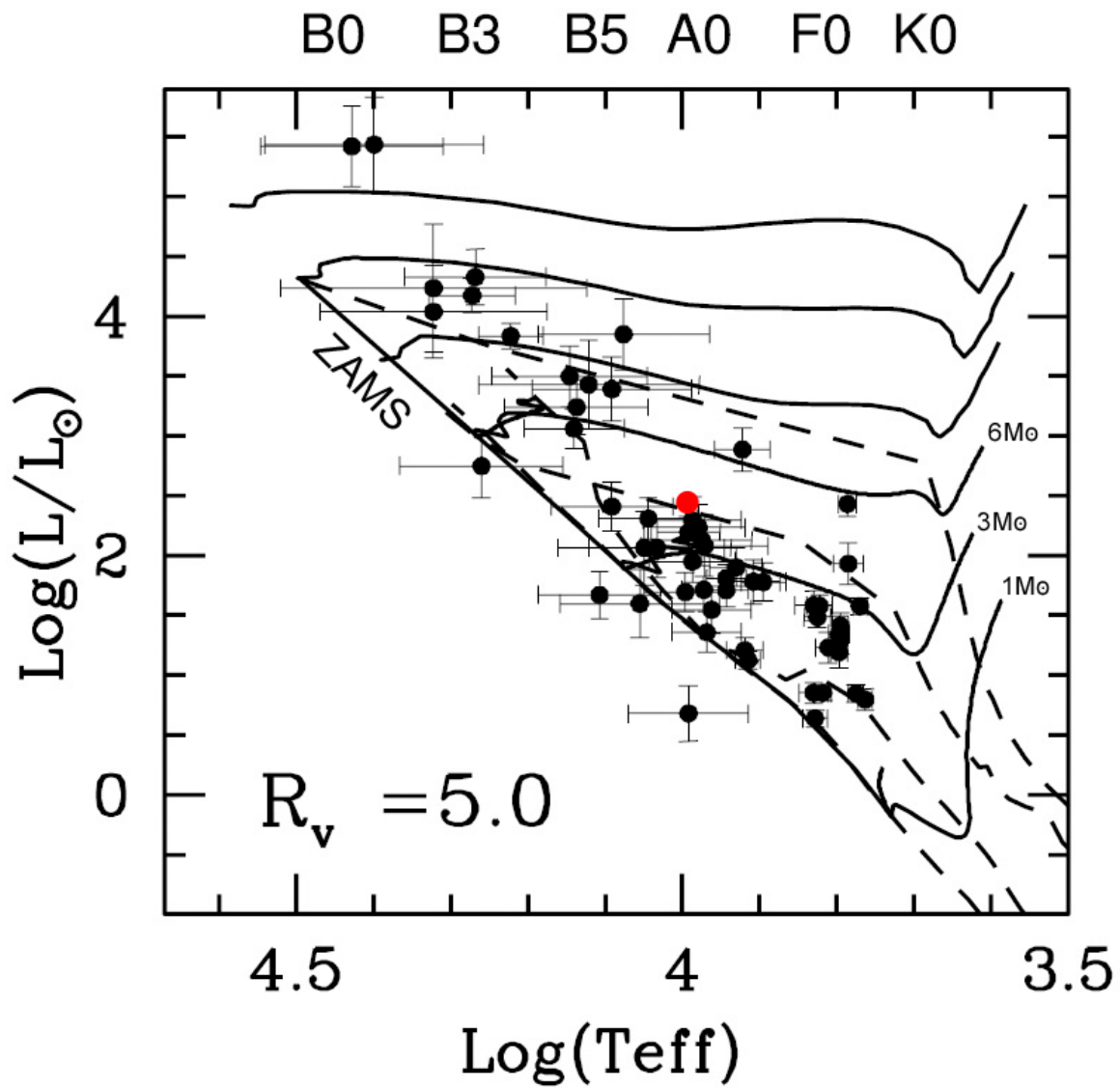

Figure 8. The position of RR Tau in the H-R diagram is shown in red among other HAEBE stars in black. Also shown are solid lines representing the evolutionary tracks of stars of different mass contracting to the Zero Age Main Sequence (ZAMS). These indicate the mass of RR Tau is around 4-5 $M_\odot$. Adapted from Figure 10 of Hernandez et al. (11).